\documentclass[twocolumn,showpacs,preprintnumbers,amsmath,amssymb]{revtex4}

\usepackage{graphicx}
\usepackage{dcolumn}
\usepackage{bm}
\usepackage{color} 

\begin{document}


\title{Towards Einstein-Podolsky-Rosen quantum channel multiplexing}

\author{Boris Hage, Aiko Samblowski and Roman Schnabel}
\address{Institut f\"ur Gravitationsphysik, Leibniz Universit\"at Hannover and Max-Planck-Institut f\"ur
Gravitationsphysik (Albert-Einstein-Institut), Callinstr.~38, 30167~Hannover, Germany}
 \email{Roman.Schnabel@aei.mpg.de}

\date{\today}

\begin{abstract}
A single broadband squeezed field constitutes a quantum communication resource that is sufficient for the
realization of a large number $N$ of quantum channels based on distributed Einstein-Podolsky-Rosen (EPR)
entangled states. Each channel can serve as a resource for, e.g. independent quantum key distribution or
teleportation protocols. $N$-fold channel multiplexing can be realized by accessing $2N$ squeezed modes at
different Fourier frequencies. We report on the experimental implementation of the $N=1$ case through the
interference of two squeezed states, extracted from a single broadband squeezed field, and demonstrate all
techniques required for multiplexing ($N>1$). 
Quantum channel frequency multiplexing can be used to optimize the exploitation of a broadband squeezed field in a quantum information task. For instance, it is useful if the bandwidth of the squeezed field is larger than the bandwidth of the homodyne detectors. This is currently a typical situation in many experiments with squeezed and two-mode squeezed entangled light.
\end{abstract}

%
%
%
%
%
%
%
%

\maketitle

\section{Introduction}
EPR entangled \cite{EPR35} optical states can be used to constitute quantum communication channels between
two distant parties. Such channels have been successfully de\-mon\-strated in both complementary regimes of
light. Entanglement in the degrees of freedom of photons can be produced by parametric down-conversion and
conditional single photon detection \cite{OMa88,KMWZS95}. This is the \emph{discrete variable} regime, in
which, more generally, arbitrary photon number states with conditional or unconditional detection might be
involved. Entanglement in the degrees of freedom of waves, i.e.~the field quadratures, provide quantum
correlations in variables possessing a continuous measurement spectrum \cite{Reid89,OPKP92,SLWKKK01,BSLR03,DHFFS07}.
In both regimes, applications of entangled states in quantum teleportation
\cite{TeleDV1,TeleDV2,FSBFKP98,BTBSRBSL03} and quantum key distribution
\cite{CrypDV1,CrypDV2,CrypCV1,CrypCV2} have attracted much attention.
\emph{Continuous variable} (CV) quantum communication is in direct analogy to conventional communication
schemes in which information is encoded in amplitude modulations (AM) and phase (frequency) modulations (FM)
of a, possibly continuous, carrier wave. The amount of quantum information  that can be transmitted, for
example in order to generate a secret key for quantum cryptography, is proportional to the bandwidth used. 
However, the useable entanglement bandwidth of a channel is typically not limited by the CV entangled field itself but rather by the speed of the high quantum efficiency homodyne detectors. Recently, Mehmet \textit{et al.} \cite{Mehmet11dB} demonstrated a broadband squeezed field with a nonclassical noise supression of up to 11.5\,dB and a bandwidth of as large as 170\,MHz. However, for Fourier frequencies above a couple of tens of MHz strong squeezing could only be observed after the dark noise of the balanced homodyne detector was subtracted. EPR quantum channel frequency multiplexing is a tool to overcome the detection bandwidth limitation.

Schori \emph{et al.}~\cite{schori:033802} experimentally demonstrated that EPR-entanglement can be produced from two frequency modes of a squeezed field. In their experiment two \textit{narrowband} longitudinal cavity modes of an optical parametric oscillator were separated with filter cavities and their correlations were detected with two balanced homodyne detectors that used frequency shifted local oscillators. Later Zhang \cite{Zha03} proposed to split a single \textit{broadband} squeezed cavity mode into $N$ pairs of upper and lower single sideband fields and to demonstrate $N$ independent EPR entangled modulation fields. 

\begin{figure}
\begin{minipage}{8.6cm}
\includegraphics[width=8.6cm]{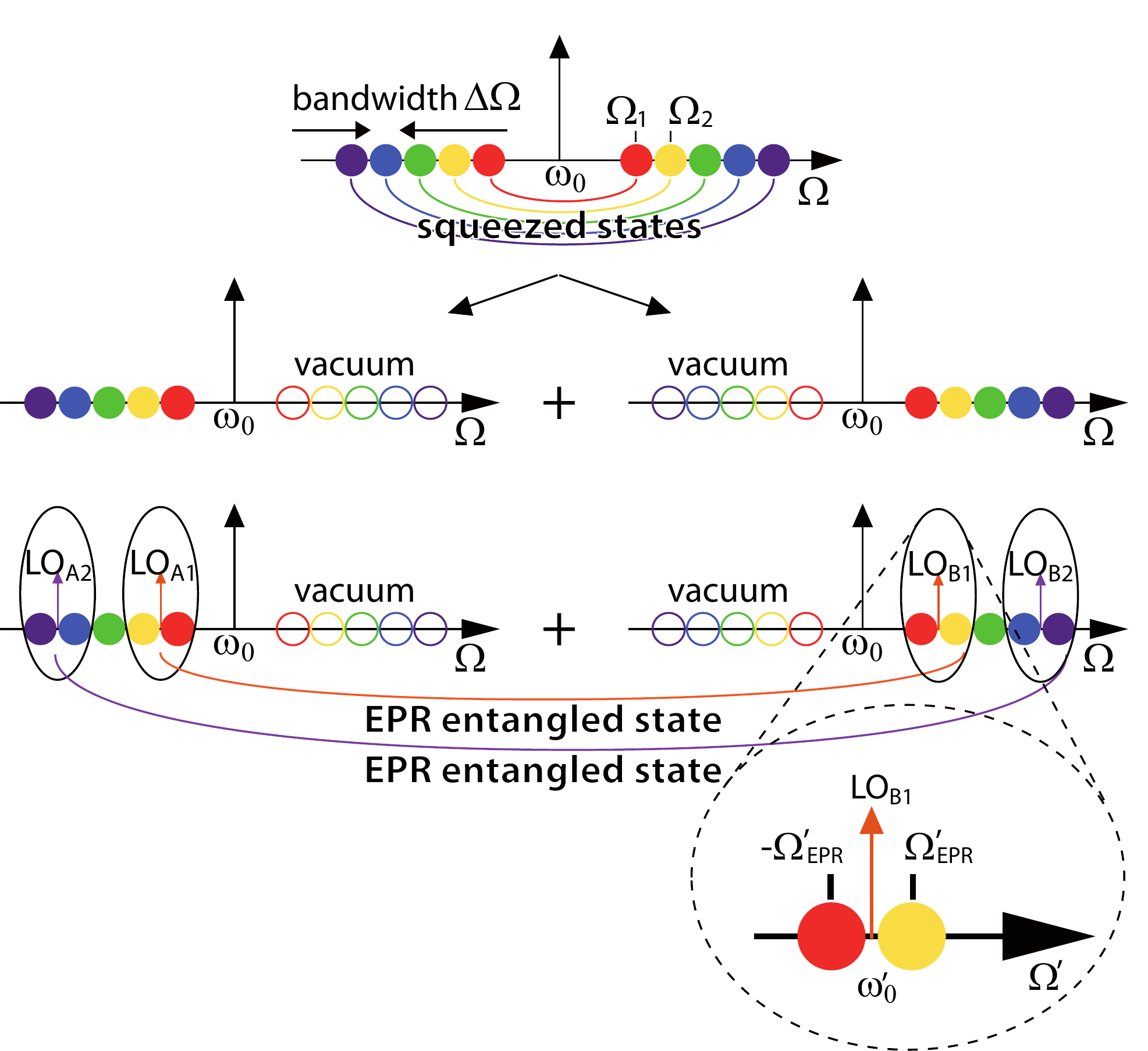}
\end{minipage}
\caption{(Colour online) Scheme for the generation of EPR
entanglement from a single broadband squeezed field in the rotating frame ($\omega_0=0$). Top: Squeezed states
in Fourier frequency space described as the sum of quantum
correlated upper and lower sideband pairs with resolution bandwidths
corresponding to the circles' diameters. Centre Line: Splitting of
the squeezed states by a frequency beam splitter, i.e.~filter
cavity. Bottom: Adding pairs of frequency shifted local oscillators
provide EPR entanglement due to interference of the squeezed states
at two different Fourier frequency. Inset: Fourier frequencies defined in the system $\omega_0^\prime=0$.} \label{sbpic}
\end{figure}

In this paper we report on the experimental generation of an EPR quantum channel from a broadband squeezed field, which corresponds to the $N=1$ case as proposed in \cite{Zha03}.
In our experiment the complete set of building blocks required for future continuous variable EPR
quantum channel multiplexing ($N>1$) is demonstrated. A multiplication of solely the classical resources of our
experiment will allow the establishment of a linearly increasing number of EPR quantum channels between pairs
of distant parties from a single broadband squeezed field.

\begin{figure}
\begin{minipage}{8.6cm}
\includegraphics[width=8.0cm]{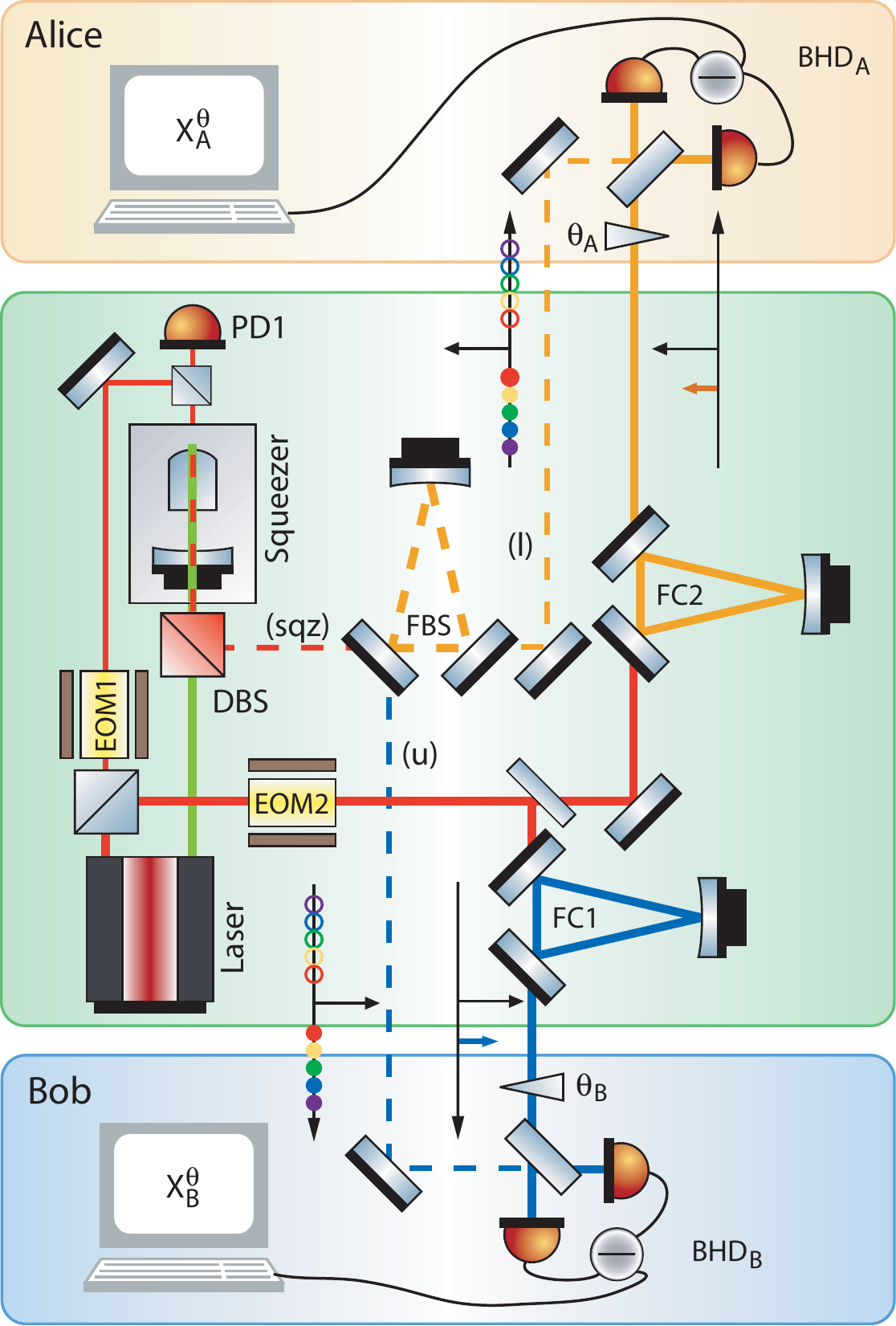}
\end{minipage}
\caption{(Colour online) Experimental setup to establish an EPR
quantum channel ($N=1$) from a single squeezed beam (sqz). The triangular
filter cavity FBS  was used as a frequency beam splitter to separate
the upper (u, blue) from the lower (l, orange) sidebands of the
nonclassical field. The electro-optical modulator EOM1 and photo
diode PD1 were used for controlling the OPA cavity length and the
pump field phase. DBS was a dichroic beam splitter, which reflected
1064\,nm and transmitted 532\,nm radiation. EOM2 generated bright
sidebands by a deep phase modulation at 7\,MHz. These sidebands
served as LOs in Alice's and Bob's homodyne detectors (BHDs). FC1
was tuned to transmit the upper sideband at +7\,MHz, FC2 transmits
the lower one at -7MHz.}
 \label{Fig2:complex}
\end{figure}

\section{Demonstration of the EPR-Paradox}

In previous continuous wave experiments CV EPR entanglement has been efficiently produced by either
type\,II optical parametric amplification (OPA) \cite{OPKP92,LCKTF05,su:070502,villar:243603} or by the
interference of two squeezed outputs from two type\,I OPA processes on a 50:50 beam splitter \cite{FSBFKP98,BSLR03,BTBSRBSL03,DHFFS07,hidehiro:110503}. Quite generally, bipartite CV, Gaussian entangled states can be represented
by the two spatial output modes of a 50:50 beam splitter, as the result of the interference of two squeezed
input modes. The states under consideration are sideband modulation fields at frequency $\Omega_{\textrm{EPR}}$ with bandwidth $\Delta\Omega$ carried by an optical field of frequency $\omega_0$, and are
formally described in the rotating frame 
by non-commuting pairs of time-dependent quadrature
operators $\hat X(\Omega_{\textrm{EPR}},\Delta\Omega,t)$ and $\hat X^\perp(\Omega_{\textrm{EPR}},\Delta\Omega,t)$, respectively, with $\frac{\Delta\Omega}{2}<\Omega_{\textrm{EPR}}\ll\omega_0$. In most CV EPR entanglement experiments so far \textit{two} optical frequencies ($\omega_0\pm\Omega_{\textrm{EPR}}$) were involved. 

In this work we used a {single} \textit{broadband} squeezed field to realize EPR entanglement. 
In this case, in total \textit{four} different optical frequencies contributed to the generation of an EPR entangled state, as shown for the
two entangled states in the lower pictures of Fig.\ref{sbpic}. For this reason such an EPR state was named a
\emph{four-mode squeezed state}  in \cite{SPSL87}. However, as in other bipartite quadrature entangled
states, the entanglement is observed between two spatial modes at a single modulation frequency
$\Omega_{\textrm{EPR}}^\prime$. The distinct feature is that here, $\Omega_{\textrm{EPR}}^\prime$ is defined with respect to two local oscillators having different optical frequencies. The top picture of Fig.~\ref{sbpic} shows that modulation fields being in
squeezed states can be described as the beats between pairs of quantum correlated upper and lower optical
sideband fields at frequencies $\omega_0\pm\Omega_i$ \cite{BachorRalph2004}. For this reason a squeezed state (of a modulation
field) has sometimes been called a \emph{two-mode squeezed state} \cite{SCa85}. We note that
a Gaussian (bipartite) EPR entangled state has also been called a \emph{two-mode squeezed state} in order to
pinpoint the presence of quantum correlations in the two \emph{spatial} modes.

In our experiment a single broadband squeezed field at 1064\,nm was generated in a
half-monolithic (hemilithic), single-ended standing wave nonlinear
cavity using type\,I OPA \cite{CVDS07}. The nonlinear medium inside the cavity was a 7\%
magnesium oxide doped lithium niobate (7\%\,MgO:LiNbO$_{3}$) crystal which
was pumped by 65\,mW of continuous wave laser radiation at 532\,nm.
The effective length of the cavity was 39\,mm and the coupler
reflectivity was $r^2=95.7\%$. The squeezing strength observed was
approximately 5.5\,dB for Fourier frequencies from 4\,MHz to
10\,MHz. At higher frequencies the squeezing strength degraded due
to the finite bandwidth of the OPA cavity, which was 25\,MHz.
At lower frequencies classical noise from control beams that sensed the OPA cavity length and the orientation of the squeezing ellipse degraded the squeezing strength. This noise may be significantly reduced by appropriate control schemes \cite{VCHFDS06}.

The generation of EPR entanglement from a single broadband squeezed field requires the implementation of the interference of two squeezed states defined at different Fourier frequencies. In order to achieve this, altogether three triangular travelling wave filter cavities were employed (FBS, FC1, and FC2), see Fig.\,\ref{Fig2:complex}. All three filter cavities consisted
of three dielectrically coated low-loss mirrors. The two plane input/output coupling mirrors had a power
transmission of $T=8500\textrm{ ppm}$ for p- and $T=300\textrm{ ppm}$ for s-polarized light. The curved
cavity end mirror showed a transmission of $T=5\textrm{ ppm}$. This provided finesse values of
$\mathcal{F}_{\textrm{p}}=370$ for p- and $\mathcal{F}_{\textrm{s}}=10500$ for s-polarization and linewidths
of $1.5\textrm{ MHz}$ and $55\textrm{ kHz}$, respectively, in accordance to the round trip length of
$52\textrm{ cm}$. The resonators were almost lossless and transmitted more than $95\%$ of resonant light
power. The lengths of the cavities were controlled via piezo electric transducers.
The filter cavity FBS was used as a \emph{frequency beam splitter}
that spatially separated the upper and lower sideband components of the
broadband squeezed field (Fig.\,\ref{Fig2:complex}). The filter cavity FBS was operated in
its low-finesse mode and was detuned by  $-7\textrm{\,MHz}$ with
respect to the carrier field at 1064\,nm. Hence it transmitted the
fields around $-7\textrm{\,MHz}$ (lower sidebands) and reflected the
rest, particularly the upper sidebands around $+7\textrm{\,MHz}$.
The upper sidebands were sent to Bob's balanced homodyne detector (BHD)
with a local oscillator (LO), which was frequency shifted by
$+7\textrm{\,MHz}$ but nevertheless had a constant phase with
respect to the main carrier field. The lower sidebands at Alice's
site were detected with a LO at $-7\textrm{\,MHz}$. We note that the
splitting of upper and lower sidebands of a broadband squeezed field
has been demonstrated before \cite{HMRRGALL05}.
The LOs for Alice's and Bob's BHDs were generated by
electro-optic phase modulation of a part of the carrier field (EOM2
in Fig.\,\ref{Fig2:complex}) and subsequent filtering (FC1 and FC2).
The modulation frequency was $7\textrm{\,MHz}$. About one third of
the power of the carrier was transferred into sidebands at
$\pm7\textrm{\,MHz}$. The modulated beam was then split into two by
a 50:50 power beam splitter. Each of these was sent to an optical
filter cavity, FC1 and FC2, respectively. Both cavities were
operated in high-finesse mode. Again, these resonators were detuned
to $+7\textrm{\,MHz}$ and $-7\textrm{\,MHz}$, respectively, and
hence transmitted only the corresponding sideband. The power of the
carrier was suppressed by a factor of $10^{5}$ which was
sufficiently high to measure quadrature operators in the frequency
shifted reference frames at Alice's and Bob's site.
Both BHDs could be phase locked to arbitrary
quadrature angles. The error signals for these control loops were
derived from the beat between the LOs and week carrier fields
co-propagating with the correlated sideband fields. In particular,
the control loops allowed the subsequent measurement of orthogonal
quadrature phases.
Both BHD signals were demodulated at 200\,kHz, low pass filtered at
50\,kHz and fed into a data acquisition system. The calculation of
the variances of each signal, the variance of the sum or difference
and covariances of the two signals was conducted by PC software.
Electronic noise of the measurement and data acquisition devices
were at least a factor of ten smaller than quadrature signals and
needed not to be taken into account.

In order to witness the presence of entanglement in our experiment
we followed \cite{BSLR03, BSLR04} and applied the inseparability
criterion introduced by Duan \emph{et al.}~\cite{DGCZ00} and the
EPR criterion introduced by Reid and Drummond \cite{RD88}. For our
setup the inseparability criterion for the presence of entanglement
in the quadratures of two fields can be written in the following
form \cite{DGCZ00,BSLR03}
\begin{equation}
\mathcal{I}_{\textrm{Insep}}=\frac{1}{4}\left(V(\hat{X}_A-\hat{X}_B)+V(\hat{X}_A^\perp
+ \hat{X}_B^\perp)\right) < 1\,. \label{eq:insep}
\end{equation}
Here, $V$ denotes variances, with the variance of a vacuum field
normalized to unity. $\hat{X}_A$ and $\hat{X}_B$ are the fields'
quadrature phase operators at Alice's and Bob's site for which the
variance of their difference $V(\hat{X}_A-\hat{X}_B)$ is minimal.
$\hat{X}_A^\perp$ and $\hat{X}_B^\perp$ are the quadrature phase
operators orthogonal to $\hat{X}_A$ and $\hat{X}_B$, respectively.

\begin{figure}
\begin{minipage}{8.6cm}
\includegraphics[width=8.6cm]{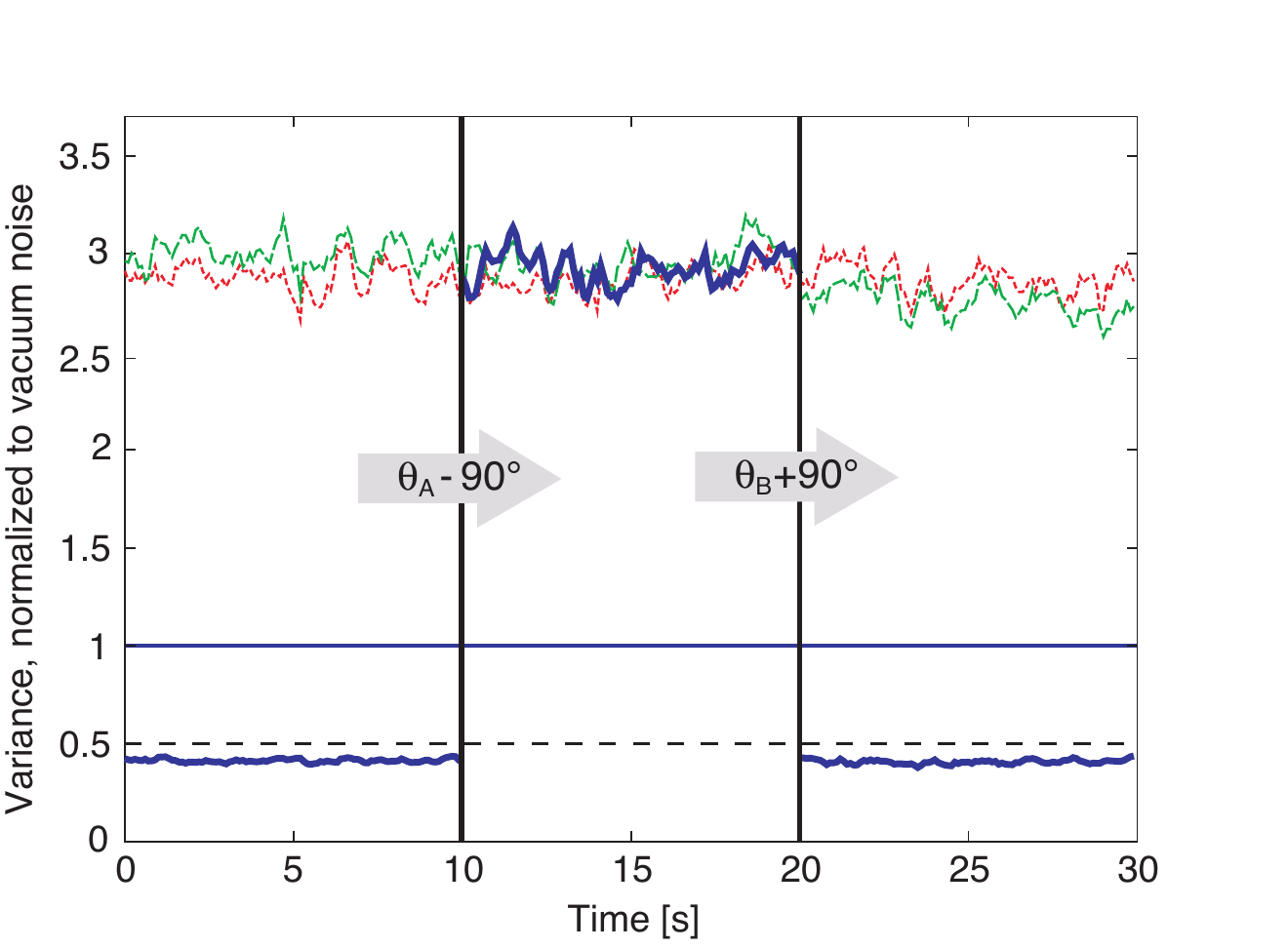}
\end{minipage}
\caption{(Colour online) Demonstration of strong entanglement between
two laser fields at Alice's and Bob's sites. The two lower
measurement traces correspond to half the variances in
Eq.\,(\ref{eq:insep}) thereby fulfilling the inequality with
$\mathcal{I}_{\textrm{Insep}}\approx0.4<1$. The dashed traces correspond to
the variances measured on the individual beams.} \label{entduan}
\end{figure}
\begin{figure*}
\begin{minipage}{17.8cm}
\includegraphics[width=16.8cm]{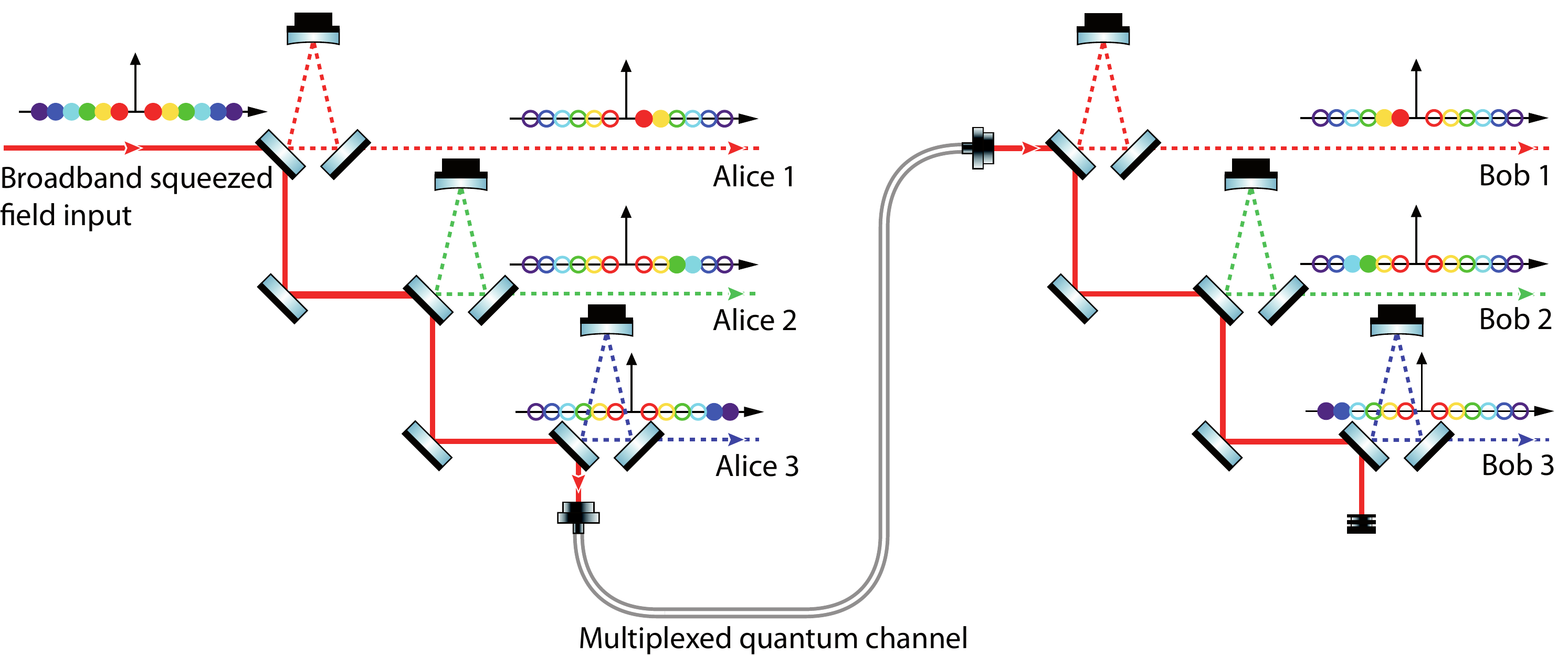}
\caption{(Colour online) Distribution of sideband fields of a squeezed field for $N=3$ EPR channel multiplexing. For each EPR channel a pair of filter cavities and a pair of frequency shifted local oscillators (not shown) are required.} 
\label{multiplexing}
\vspace{3mm}
\end{minipage}
\end{figure*}

Fig.\,\ref{entduan} presents consecutive measurement time series of
$V(\hat{X}_A)$, $V(\hat{X}_B)$, and $1/2\cdot
V(\hat{X}_A-\hat{X}_B)$ (left), $V(\hat{X}_A)$,
$V(-\hat{X}_B^\perp)$, and $1/2\cdot V(\hat{X}_A+\hat{X}_B^\perp)$ (centre),
as well as $V(\hat{X}_A^\perp)$, $V(-\hat{X}_B^\perp)$, and
$1/2\cdot V(\hat{X}_A^\perp+\hat{X}_B^\perp)$ (right). During the
measurement time shown the BHDs at Alice's and Bob's
site were phase controlled, and were quickly and sequentially
switched from a $\hat{X}$ to a $\hat{X}^\perp$ measurement.
Additionally, the vacuum noise levels of the detectors were measured
and used to normalize the traces shown. Using Eq.\,(\ref{eq:insep}),
the data in Fig.\,\ref{entduan} clearly demonstrates the presence of
entanglement with $\mathcal{I}_{\textrm{Insep}}=0.41\pm0.02$. This value not
only fulfils the inequality, but is also smaller than 0.5 proving
that less than a full unit of vacuum noise entered the detection of
entanglement in our setup. In this case the entanglement is strong
enough to observe the Einstein-Podolsky-Rosen paradox. By optimizing the gain factors on subsequent measurement results of two non-commuting quadratures on Bob's field we were able to infer the corresponding results on Alice's field more precisely than suggested by the existence of vacuum noise. The EPR paradox is observed if the following
EPR criterion is fulfilled \cite{RD88}:
\begin{eqnarray}
    \label{VcvVcv}\nonumber
\mathcal{E}_{\textrm{EPR}} \!  = \! \min_{g}{\left \langle \left
(\delta \hat X_{A}  -  g \,\delta \hat X_{B}  \right )^{2} \right
\rangle}\,\;\;\;\;\;\;\;\;\;\;\;\;\;\;\;\;\;\;\;\;\;\;\;\;\;  \\
\! \times \,\;  \! \min_{g^{\perp}}{\left \langle \left (\delta \hat X^{\perp}_{A}  
+  g^{\perp} \delta \hat X^{\perp}_{B}  \right )^{2} \right \rangle} <1\,,
\end{eqnarray}

with $\delta \hat X=\hat X \!-\! \langle \hat X \rangle$ and $g$ and $g^{\perp}$ being parameters that are experimentally adjusted
to minimize the two expectation values in Eq.\,(\ref{VcvVcv}). We
observed a value of $\mathcal{E}_{\textrm{EPR}} = 0.64\pm0.02$.

\section{Discussion}
In our experiment two squeezed states at Fourier frequencies of $\Omega_{1}=6.8$\,MHz and
$\Omega_{2}=7.2$\,MHz with bandwidths of $\Delta\Omega=2\times 50$\,kHz were brought to interference in order
to produce an EPR entangled state at the Fourier frequency of $\Omega_{\textrm{EPR}}^\prime=200$\,kHz with respect to the frequency
shifted local oscillators. The initial squeezed states were carried by a single broadband squeezed field. The
same field carried more squeezed states which were not used by our quantum channel. Fig.\,\ref{multiplexing}
shows how a single broadband squeezed field can be used to provide the nonclassical resource of three ($N$=3
multiplexed) EPR quantum channels. Each channel can be used for quantum communication tasks between Alice and
Bob, or alternatively can be established between different senders and receivers.
Quantum channel frequency multiplexing can be used if the individual quantum
communication tasks require less bandwidth than provided by the
entangled light source. 
However, a more obvious application is to overcome the bandwidth limitations set by balanced homodyne detectors. 
In principle the bandwidth of a squeezed light source design is just limited by the available second harmonic pump power and by the phase
matching bandwidth of the nonlinear material used. For periodically poled materials such as PPKTP, the phase matching bandwidth is of the order of a nanometre and can therefore cover hundreds of GHz \cite{Boyd2003}. High quantum efficiency BHDs with low electronic moise used in nonclassical light applications typically have detection bandwidths of just several tens of MHz, because faster BHDs require smaller photo diodes, and consequently, optical local oscillators with less powers in order to avoid too high thermal loads, see for example
\cite{Hobbs2000}. A significant increase of bandwidth is certainly possible, however, the optical bandwidths of
nonclassical light sources will probably not be reached. Multiplexing of the
nonclassical frequency band can solve this gap. $N$ pairs of
frequency shifted LOs picked from a frequency comb with frequency
separation of twice the electronic detection bandwidth can
complement the scheme shown in  Fig.\,\ref{multiplexing}. Thus
high speed quantum communication with $N$-times the electronic
detection bandwidth can be achieved.

\section{Summary}
To summarize, we experimentally demonstrated that a single broadband squeezed field can be used to establish an EPR quantum channel. Additional EPR quantum channels can be produced by increasing the classical resources of our experiment without increasing its nonclassical resources. The EPR quantum channel multiplexing discussed will allow an efficient use of broadband nonclassical fields for the realization of ultrahigh quantum information transmission rates.

\begin{acknowledgments}
We would like to acknowledge financial support by the Deutsche Forschungsgemeinschaft through the SFB407. We also thank U.~Andersen, R.~Fillip, J.~Fiur\'{a}\v{s}ek, N.~Lastzka, and G.~Leuchs for many helpful discussions regarding entanglement of sideband
fields and J.~DiGuglielmo,  A.~Franzen and H.~Vahlbruch regarding the control of entangled quadrature phases.
\end{acknowledgments}


\end{document}